# Unidirectional lasing via vacuum induced coherent in defective atomic lattice

Xinfu Zheng,[1] Chen Peng,[1] Duanfu Chen,[1] Yiting Zheng,[1]
Hanxiao Zhang,[1] Dong Yan,[1] Jinhui Wu[†],[2] and Hong Yang[*1]

[1]*School of Physics and Electronic Engineering, Hainan Normal University, Haikou 571158, People's Republic of China*
[2]*Center for Quantum Sciences and School of Physics,
Northeast Normal University, Changchun 130024, People's Republic of China*

(Dated: August 20, 2025)

We skillfully utilized vacuum induced coherence to amplify the probe light, and then successfully achieved both nonreciprocal reflection and lasing oscillation in a single physical system by leveraging the distributed feedback and spatial symmetry breaking effect of the one-dimensional defective atomic lattice. This innovative scheme for realizing unidirectional reflection lasing (URL) is based on both non-Hermitian degeneracy and spectral singularity (NHDSS, means $\lambda_+^{-1} \simeq \lambda_-^{-1} \to 0$). Therefore, we analyze the modulation of parameters such as the lattice structure and external optical fields in this system to find NHDSS point, and further verified the conditions for its occurrence by solving the transcendental equation of susceptibility satisfying the NHDSS point, as well as analyzed its physical essence. Our mechanism is not only beneficial for the integration of photonic devices in quantum networks, but also greatly improves the efficiency of optical information transmission.

PACS numbers: 64.70.Tg, 03.67.-a, 03.65.Ud, 75.10.Jm

## I. INTRODUCTION

Lasers have important applications in medical, military, and information technology fields due to their special characteristics such as directionality and high intensity [1–4]. However, due to bidirectional optical field oscillation, reflected light can cause laser frequency shift and increased phase noise, so traditional lasers typically requiring external isolators. The intrinsic unidirectional lasing effect of non-reciprocal lasers precisely overcomes this limitation, which provides a new path for the miniaturization and high reliability of photonic systems, especially demonstrating irreplaceable value in cutting-edge fields such as integrated photonics and quantum technologies [5–8]. In recent years, non-reciprocal lasers have been studied in many different physical systems. Such as Brillouin lasers [9, 10], topological lasers [11–13], phonon lasers [14–16], magon lasers [17–19], and circularly polarized lasers [20, 21].

In the atom-light intercation system, the generation of non-reciprocal lasing is dependent both on the establishment of strong oscillations and distributed feedback of light field assisted by gain medium, and on a non-reciprocal mechanism [22–25]. Some typical gain systems in atomic media have been proposed and studied in light amplification, e.g., the coherent gain atomic system that forms closed-loop transitions under the action of a microwave field can enable the probe field to be amplified at the electromagnetically induced transparency (EIT) window [38]; and the probe light gain systems induced by the nonlinear effect of four-wave mixing (FWM) [26–28], or spontaneously generated coherence (SGC) [29–31]. In particular, the SGC is also called vacuum induced coherence (VIC), which has been recently observed in experiments [32–36]. In addition, based on FWM gain, lasing emission has been achieved with cold $^{87}$Rb atoms trapped in the optical lattice [39, 40]. Moreover, in optical lattices, lasing has been realized using a cold atomic ensemble of $^{88}$Sr based on atomic clock principles [41]. This is because atomic lattices can replace resonant cavities to effectively realize distributed feedback lasing oscillation. However, non-reciprocal lasing in atomic lattice systems has not been studied yet. In our recent work, the non-reciprocal or even the unidirectional reflection has been achieved in a defective atomic lattice based on the symmetry breaking effect [42], and further realized unidirectional reflection amplification under low-gain in this scheme with establishing coherent gain atomic system by microwave field coupling [43].

Next, in order to achieve the non-reciprocity and lasing oscillation in a single system, which is well for integrated and high-efficiency photonic devices, we utilizing the coherent driven three-level $V$-type active atomic system for amplifying a weak probe field via VIC [see Fig. 1(a)], and the one-dimensional (1D) defective atomic lattice to break the spatial symmetry of probe susceptibility and provide distributed feedback regime [see Fig. 1(c)]. The tunable unidirectional reflection lasing (URL) can be realized by setting parameters appropriately, that corresponding to the non-Hermitian degeneracy (NHD) point ($\lambda_+^{-1} \simeq \lambda_-^{-1}$ indicates the unidirectional reflection) [44, 45] and the spectral singularity (SS) point ($\lambda_+^{-1} \to 0$ means generating lasing output) [46–50]. Under the condition that the above two requirements are satisfied, it actually means that $\lambda_+^{-1} \simeq \lambda_-^{-1} \to 0$, we define it as NHDSS-URL. Furthermore, we try to find the NHDSS point by adjusting the lattice structure parameters, the Rabi frequency and detuning of the coupling field, the closed-loop phase $\phi_F$, and the angle of the dipole moments. In order to verify the conditions and physical essence for realizing URL corresponding to the NHDSS point, we solve the transcendental equation of susceptibility that satisfies this point. Meanwhile, we also analyse the probe susceptibility corresponding to unidirec-tional reflection (UR), unidirectional reflection amplification (URA), and the URL respectively.

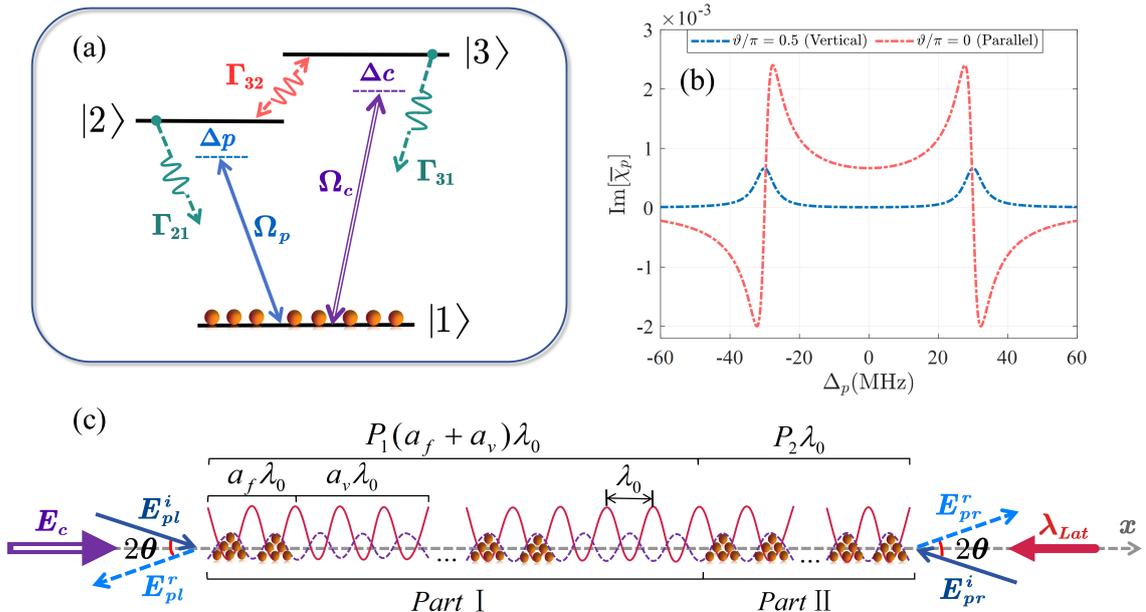

FIG. 1: (a) Energy level diagram of a closed-loop three-level $V$-type atomic system driven by a weak probe field, and a strong coupling fields with the help of cross-damping rate $\Gamma_{32}$ (i.e., VIC). (b) The imaginary part of average probe susceptibility Im$[\overline{\chi}_p]$ $v.s.$ probe detuning $\Delta_p$ with $\vartheta/\pi = 0.5$ (blue dashed line) and $\vartheta/\pi = 0$ (red dashed line) in $\Omega_{c_0} = 30$ MHz, $\Delta_c = 0$, $\phi_F/\pi = 0$. (c) The 1D defective atomic lattice with the width $\lambda_0$ for each period, which comprises Part I and Part II. The probe filed $\mathbf{E}_p$ is incident from either the left side (denoted by $\mathbf{E}_{pl}^i$) or the right side (denoted by $\mathbf{E}_{pr}^i$) with a small incident angle $\theta$ relative to $x$-axis. The relevant reflected fields are denoted by $\mathbf{E}_{pl}^r$ and $\mathbf{E}_{pr}^r$. The parameters are $\Omega_{p_0} = 0.5$ MHz, $\Gamma_{31} = 6.1$ MHz, $\Gamma_{21} = 5/9 \times 6.1$ MHz, $\Gamma_{32} = \sqrt{5/9} \times 6.1 \times \cos\vartheta$ MHz, $\lambda_{Lat} = 781.3$ nm, $\lambda_{21} = 780.24$ nm, $\Delta\lambda_{Lat} = 0.9$ nm, $\eta = 5.0$, $N_0 = 7 \times 10^{10}$cm$^{-3}$, $\mathbf{d}_{12} = 1.0907 \times 10^{-29}$ C·m.

## II. THEORETICAL MODEL AND EQUATIONS

As shown in Fig. 1(a), the cold $^{85}$Rb atoms are driven into a three-level $V$-type system, by a weak probe field with frequency $\omega_p$ (amplitude $\mathbf{E}_p$) and strong coupling field with frequency $\omega_c$ (amplitudes $\mathbf{E}_c$), forming a close-loop based on vacuum-induced coherent (VIC) trapped in a defective atomic lattice. The Rabi frequency (detuning) of probe is denoted by $\Omega_p = \mathbf{E}_p \cdot \mathbf{d}_{12}/2\hbar$ ($\Delta_p = \omega_{21} - \omega_p$) interact via the dipole-allowed transition $|1\rangle \leftrightarrow |2\rangle$. The Rabi frequency (detuning) of coupling field $\Omega_c = \mathbf{E}_c \cdot \mathbf{d}_{13}/2\hbar$ ($\Delta_c = \omega_{31} - \omega_c$) drives the dipole-allowed transition $|1\rangle \leftrightarrow |3\rangle$. The matrix element $\mathbf{d}_{ij} = \langle i|\mathbf{d}|j\rangle$ is used to denote the dipole moment of transition $|i\rangle$ to $|j\rangle$. Specifically, the energy levels $|1\rangle$, $|2\rangle$, and $|3\rangle$ refer to $|5S_{1/2}, F = 3\rangle$, $|5P_{3/2}, F = 3\rangle$, and $|5P_{3/2}, F = 4\rangle$, respectively, in the D2 line of $^{85}$Rb atoms. $\Gamma_{32}$ represents the cross-damping rate between the excited states $|3\rangle$ and $|2\rangle$, namely, the VIC as mentioned in work of Han et al. [32], whereas $\Gamma_{31}$ and $\Gamma_{21}$ denote the spontaneous decay rates the atoms transition from level $|3\rangle$ and $|2\rangle$ to level $|1\rangle$, respectively. Notably, $\Gamma_{32} = \sqrt{\Gamma_{31}\Gamma_{21}} \cos\vartheta$ (i.e., VIC), here $\vartheta$ is the angle between the two dipole moments with $\vartheta/\pi \in [0, 0.5]$, corresponding to the maximum VIC and without VIC. and the damping rate of the excited state $|3\rangle$ to the ground state $|1\rangle$ is $\Gamma_{31} = 6.1$ MHz, whereas the damping rate of the excited state $|2\rangle$ to the ground state $|1\rangle$ is $\Gamma_{21} = 5/9 \times 6.1$ MHz because the excited state $|2\rangle$ decays to the ground state $|1\rangle$ only fractionally with the branching ratio of 5/9 [32, 33, 37]. In Fig. 1(c), a 1D defective atomic lattice engineered into two periodic structural parts: part I with period $P_1$, consisting of $a_f$ filled lattice cells and $a_v$ vacant lattice cells; part II with period $P_2$ of filled lattice cells. It should be emphasized that all atoms trapped in each filled lattice cell exhibit a Gaussian distribution. The probe light is incident from both sides of the 1D defective atomic lattice with a small angle $\theta$, and its left/right reflectivities can be solved via numerical simulation.

Within the electric-dipole and rotating-wave approximations, the total Hamiltonian of the system which describes the atom-field coupling can be expressed as

$$\hat{H}_{tot} = \omega_{21}\hat{\sigma}_{22} + \omega_{31}\hat{\sigma}_{33} \\ - (\Omega_p e^{-i\omega_p t}\hat{\sigma}_{21} + \Omega_c e^{-i\omega_c t}\hat{\sigma}_{31} + \text{H.c.}), \quad (1)$$

where the corresponding Rabi frequencies are redefined as $\Omega_p = \Omega_{p_0} e^{i\phi_p}$ and $\Omega_c = \Omega_{c_0} e^{i\phi_c}$, where $\Omega_{p_0}$ and $\Omega_{c_0}$ are real numbers. Then, by choosing the external classical optical field frequency, we can perform a unitary transformation to a rotating coordinate frame described by the unitary operator $\hat{U} = e^{-i\hat{H}_f t}$, where $\hat{H}_f = \omega_p \hat{\sigma}_{22} + \omega_c \hat{\sigma}_{33}$. In terms of the formula $\hat{H}_{rot} = \hat{U}^\dagger \hat{H}_{tot} \hat{U} - i\hat{U}^\dagger(\partial \hat{U}/\partial t)$, we obtain time-independent Hamiltonian, namely

$$\hat{H}_{rot} = \Delta_p \hat{\sigma}_{22} + \Delta_c \hat{\sigma}_{33} - (\Omega_p \hat{\sigma}_{21} + \Omega_c \hat{\sigma}_{31} + \text{H.c.}), \quad (2)$$

without loss of generality, we adopt the same method as in Ref. [34, 35], thus the atomic transformations are redefined as $\hat{\sigma}_{21} \to \hat{\sigma}_{21} e^{-i\phi_p}$ and $\hat{\sigma}_{31} \to \hat{\sigma}_{31} e^{-i\phi_c}$, where the total closed-loop phase $\phi_F = \phi_c - \phi_p$ playing a crucial role in finely modulating unidirectional lasing in this paper. Based on this transformation and considering the incoherent (dissipative) processes, the density-matrix operator $\hat{\rho}$ of the atomic system is described by the Lindblad master equation in the Born-Markov approximation:

$$\begin{aligned}
\partial_t \hat{\rho} = &-i\left[\hat{H}_{tra}, \hat{\rho}\right] + \Gamma_{21} \mathcal{L}(\hat{\sigma}_{12}) \hat{\rho} + \Gamma_{31} \mathcal{L}(\hat{\sigma}_{13}) \hat{\rho} \\
&+ \Gamma_{32} e^{i\phi_F} \left(\hat{\sigma}_{13} \hat{\rho} \hat{\sigma}_{12}^\dagger - \hat{\sigma}_{12}^\dagger \hat{\sigma}_{13} \hat{\rho}/2 - \hat{\rho} \hat{\sigma}_{12}^\dagger \hat{\sigma}_{13}/2\right) \\
&+ \Gamma_{32} e^{-i\phi_F} \left(\hat{\sigma}_{12} \hat{\rho} \hat{\sigma}_{13}^\dagger - \hat{\sigma}_{13}^\dagger \hat{\sigma}_{12} \hat{\rho}/2 - \hat{\rho} \hat{\sigma}_{13}^\dagger \hat{\sigma}_{12}/2\right),
\end{aligned} \quad (3)$$

with

$$\hat{H}_{tra} = \Delta_p \hat{\sigma}_{22} + \Delta_c \hat{\sigma}_{33} - (\Omega_{p_0} \hat{\sigma}_{21} + \Omega_{c_0} \hat{\sigma}_{31} + \text{H.c.}), \quad (4)$$

the above Lindblad superoperator $\mathcal{L}(\hat{\sigma})$ describes the dissipative coupling to the environment and is given by the form $\mathcal{L}(\hat{\sigma})\hat{\rho} = \hat{\sigma}\hat{\rho}\hat{\sigma}^\dagger - \hat{\sigma}^\dagger\hat{\sigma}\hat{\rho}/2 - \hat{\rho}\hat{\sigma}^\dagger\hat{\sigma}/2$. Finally, the density-matrix equations are written as follows above:

$$\begin{aligned}
\partial_t \hat{\rho}_{11} &= i\Omega_{p_0}(\hat{\rho}_{21} - \hat{\rho}_{12}) + i\Omega_{c_0}(\hat{\rho}_{31} - \hat{\rho}_{13}) + \Gamma_{21}\hat{\rho}_{22} \\
&\quad + \Gamma_{31}\hat{\rho}_{33} + \Gamma_{32}(\hat{\rho}_{32} e^{i\phi_F} + \hat{\rho}_{23} e^{-i\phi_F}), \\
\partial_t \hat{\rho}_{22} &= i\Omega_{p_0}(\hat{\rho}_{12} - \hat{\rho}_{21}) - \Gamma_{21}\hat{\rho}_{22} \\
&\quad - \Gamma_{32}(\hat{\rho}_{32} e^{i\phi_F} + \hat{\rho}_{23} e^{-i\phi_F})/2, \\
\partial_t \hat{\rho}_{21} &= i\Omega_{p_0}(\hat{\rho}_{11} - \hat{\rho}_{22}) - i\Omega_{c_0}\hat{\rho}_{23} \\
&\quad - (i\Delta_p + \Gamma_{21}/2)\hat{\rho}_{21} - \Gamma_{32} e^{i\phi_F} \hat{\rho}_{31}/2, \\
\partial_t \hat{\rho}_{31} &= i\Omega_{c_0}(\hat{\rho}_{11} - \hat{\rho}_{33}) - i\Omega_{p_0}\hat{\rho}_{32} \\
&\quad - (i\Delta_c + \Gamma_{31}/2)\hat{\rho}_{31} - \Gamma_{32} e^{-i\phi_F} \hat{\rho}_{21}/2, \\
\partial_t \hat{\rho}_{32} &= i\Omega_{c_0}\hat{\rho}_{12} - i\Omega_{p_0}\hat{\rho}_{31} - \Gamma_{32} e^{-i\phi_F}(\hat{\rho}_{22} + \hat{\rho}_{33})/2 \\
&\quad - [i(\Delta_c - \Delta_p) + (\Gamma_{31} + \Gamma_{21})/2]\hat{\rho}_{32}.
\end{aligned} \quad (5)$$

The above equations are restricted by the conjugation condition $\hat{\rho}_{ij} = \hat{\rho}_{ji}^*$ and the trace condition $\sum_i \hat{\rho}_{ii} = 1$. Under the weak probe field condition, we can obtain $\hat{\rho}_{21}$ in terms of $\hat{\rho}_{11}$ and $\hat{\rho}_{33}$, that is

$$\hat{\rho}_{21} = i\frac{X'\hat{\rho}_{11} + Y'\hat{\rho}_{33} + Z'}{\gamma'_{21}\gamma'_{31}\gamma'_{32} + |\Omega_{c_0}|^2 \gamma'_{31} - \gamma'_{32}\Gamma_{32}^2/4}, \quad (6)$$

with

$$\begin{aligned}
X' &= 2\Omega_{p_0} \gamma'_{31} \gamma'_{32} - \Omega_{c_0}(\gamma'_{32} + \gamma'_{31})\Gamma_{32} e^{i\phi_F}/2, \\
Y' &= \gamma'_{32}(\Omega_{p_0}\gamma'_{31} + \Omega_{c_0}\Gamma_{32} e^{i\phi_F}/2), \\
Z' &= \gamma'_{31}(-\Omega_{p_0}\gamma'_{32} + \Omega_{c_0}\Gamma_{32} e^{i\phi_F}/2),
\end{aligned} \quad (7)$$

and we define $\gamma'_{31} = i\Delta_c + \Gamma_{31}/2$, $\gamma'_{21} = i\Delta_p + \Gamma_{21}/2$, $\gamma'_{32} = i(\Delta_p - \Delta_c) + (\Gamma_{31} + \Gamma_{21})/2$. Clearly, VIC and total closed-loop phase $\phi_F$ has a significant impact on the gain of the atomic system in the above formula. Under the steady state $\partial_t \hat{\rho}_{ij} = 0$, we can obtain the numerical solution of $\hat{\rho}_{21}$, which is governed by the probe detuning $\Delta_p$. Then, the steady-state probe susceptibility $\chi_p$ of each filled cell can be obtained in $V$-type atomic system, given by

$$\chi_p(x) = \frac{N(x) |\mathbf{d}_{12}|^2 \hat{\rho}_{21}}{\hbar \varepsilon_0 \Omega_{p_0}}. \quad (8)$$

Here, the real and imaginary parts of susceptibility, corresponding to dispersion and absorption lines of probe beam, $\varepsilon_0$ is the dielectric constant in vacuum, and the spatial atomic density $N(x)$, which can be considered as a Gaussian distribution in each filled cell, is given by

$$N(x) = \frac{N_0 \lambda_0}{\sigma_x \sqrt{2\pi}} \cdot e^{[-(x-x_0)^2/2\sigma_x^2]}. \quad (9)$$

Where $N_0$ is the average atomic density, $x_0$ is the trap center, $\sigma_x = \lambda_{Lat}/(2\pi\sqrt{\eta})$ is the half-width with $\eta = 2U_0/(\kappa_B T)$ related to the capture depth $U_0$ and temperature $T$. $\lambda_0 = \lambda_{Lat}/2$ is the width of each cell, with $\lambda_{Lat}$ being the wavelength of a red-detuned retroreflected laser beam forming the optical lattice. For the Bragg condition, the incident angle is given by $\cos\theta = \lambda_p/\lambda_{Lat0}$, where $\lambda_{Lat0} = \lambda_{Lat} - \Delta\lambda_{Lat}$ with the geometric Bragg shift $\Delta\lambda_{Lat}$. It is worth emphasizing that the condition for trapping atoms is $\lambda_{Lat} > \lambda_{21}$, where $\lambda_{21}$ is the wavelength of the transition $|2\rangle \leftrightarrow |1\rangle$.

The reflection and transmission properties of the whole defective atomic lattice can be characterized by a $2 \times 2$ unimodular transfer matrix $M$, which relates the forward (electric) field amplitudes ($E_{pl}^i$ and $E_{pl}^r$) to the backward (electric) field amplitudes ($E_{pr}^i$ and $E_{pr}^r$). Then, we divide each filled lattice cell into a sufficient number of thin layers (e.g., $j_{\max}$ layers, $j_{\max} = 100$), which can be regarded as homogeneous media. Thus, the reflection coefficients on both sides of one layer are equal, that is, $r^l(x_j) = r^r(x_j) \equiv r(x_j)$. And with the same transmission coefficient on the left/right side $t(x_j)$, the transfer matrix $m(x_j)$ of a single layer can be written as

$$m(x_j) = \frac{1}{t(x_j)} \begin{bmatrix} t(x_j)^2 - r(x_j)^2 & r(x_j) \\ -r(x_j) & 1 \end{bmatrix}. \quad (10)$$

Eq. (10) establishes the relationship between the forward field and backward field on both sides of $j$th layer in each filled lattice cell, with $j \in [1, j_{\max}]$. This depends on the complex susceptibility $\chi_p(x_j)$, because the refractive index that determines the reflection and transmission coefficients [51] can be expressed as $n_p(x_j) = \sqrt{1 + \chi_p(x_j)}$. Therefore, the transfer matrix of one filled lattice cell can be expressed as

$$M_f = \Pi_{j=1}^{100} m(x_j). \tag{11}$$

For the refractive index $n_v = 1$ of vacant lattice cells. Thus, the transfer matrix of one vacant lattice cell is written as

$$M_v = \frac{1}{t(x)} \begin{bmatrix} t(x)^2 & 0 \\ 0 & 1 \end{bmatrix} = \begin{bmatrix} e^{ik\lambda_0} & 0 \\ 0 & e^{-ik\lambda_0} \end{bmatrix}, \tag{12}$$

here $k = 2\pi \cos\theta/\lambda_p$ is the probe wave number. Then, the total transfer matrices is represented as

$$M = (M_f)^{P_2} \cdot [(M_v)^{a_v} \cdot (M_f)^{a_f}]^{P_1}. \tag{13}$$

Correspondingly, the reflectivities on both sides of the one-dimensional defective atomic lattice can be obtained:

$$R^l = |r^l|^2 = \left| -\frac{M(2,1)}{M(2,2)} \right|^2,$$
$$R^r = |r^r|^2 = \left| \frac{M(1,2)}{M(2,2)} \right|^2. \tag{14}$$

The transfer matrix of Eq. (13) obtained by the layered method, although with certain approximations, can be fully used to solve the left and right reflectivities that do not require error correction. However, in order to provide an accurate transfer matrix and better understand the physical essence, we establish an analytical expression of the transfer matrix $M'$ based on Bloch theory [52–55], which can be used to represent the left and right reflection coefficients. The total matrix $M'$ of the medium is formed by a composite periodic structure consisting of the minimal basis units $M'_f$ and $M'_v$ (corresponding to a filled lattice cell and a vacant lattice cell, respectively.). Therefore, by analyzing the eigenvalue equations of matrix $M'_f$ and $[(M'_v)^{a_v} \cdot (M'_f)^{a_f}]$, we can find that there exist two distinct Bloch wave vectors $\mathcal{K}_1$ and $\mathcal{K}_2$. We know the incident probe field is expected to experience nonlinear Bragg scattering and the eigenvalue equation of matrix $M'_f$ and $[(M'_v)^{a_v} \cdot (M'_f)^{a_f}]$ can be obtained by

$$\begin{bmatrix} E^i_{pr} \\ E^r_{pr} \end{bmatrix} = M'_f \begin{bmatrix} E^i_{pl} \\ E^r_{pl} \end{bmatrix} = e^{i\mathcal{K}_1 \lambda_0} \begin{bmatrix} E^i_{pl} \\ E^r_{pl} \end{bmatrix} \tag{15}$$

and

$$\begin{bmatrix} E^i_{pr} \\ E^r_{pr} \end{bmatrix} = (M'_v)^{a_v} \cdot (M'_f)^{a_f} \begin{bmatrix} E^i_{pl} \\ E^r_{pl} \end{bmatrix}$$
$$= e^{i\mathcal{K}_2(a_f + a_v)\lambda_0} \begin{bmatrix} E^i_{pl} \\ E^r_{pl} \end{bmatrix}, \tag{16}$$

respectively. $\mathcal{K}_1$ corresponds to a filled lattice cell of length $\lambda_0$, which depends on matrix $M'_f$, while $\mathcal{K}_2$ corresponds to quasi-periodic lattice cells of length $(a_f + a_v)\lambda_0$, depending on matrix $[(M'_v)^{a_v} \cdot (M'_f)^{a_f}]$. Thus $\mathcal{K}_1$ and $\mathcal{K}_2$ can be obtained from Eq. (15) and Eq. (16) by analyzing the eigenvalue equation

$$\mathcal{K}_1 = \pm \frac{\cos^{-1}\left[\frac{1}{2}\text{Tr}\left\{M'_f\right\}\right]}{\lambda_0},$$
$$\mathcal{K}_2 = \pm \frac{\cos^{-1}\left[\frac{1}{2}\text{Tr}\left\{(M'_v)^{a_v} \cdot (M'_f)^{a_f}\right\}\right]}{(a_f + a_v)\lambda_0}. \tag{17}$$

We construct the matrix $M'$ of whole medium starting from the two minimal basis units $M'_f$ and $M'_v$ according to Eq. (13),

$$[(M'_v)^{a_v} \cdot (M'_f)^{a_f}]^{P_1} = -\frac{\sin[(P_1 - 1)\mathcal{K}_2(a_f + a_v)\lambda_0]}{\sin[\mathcal{K}_2(a_f + a_v)\lambda_0]} \hat{\mathbf{I}}$$
$$+ \frac{\sin[P_1\mathcal{K}_2(a_f + a_v)\lambda_0]}{\sin[\mathcal{K}_2(a_f + a_v)\lambda_0]} (M'_v)^{a_v} \cdot (M'_f)^{a_f},$$
$$(M'_f)^{P_2} = -\frac{\sin[(P_2 - 1)\mathcal{K}_1\lambda_0]}{\sin[\mathcal{K}_1\lambda_0]} \hat{\mathbf{I}} + \frac{\sin[P_2\mathcal{K}_1\lambda_0]}{\sin[\mathcal{K}_1\lambda_0]} M'_f, \tag{18}$$

where $\hat{\mathbf{I}}$ is unity matrix and $(M'_f)^{a_f}$ has exactly the same form as $(M'_f)^{P_2}$. For simplicity, we define the general formula of the two matrix coefficients on the right-hand side of Eq. (18) using $\widetilde{\mathcal{N}}$ and $\widehat{\mathcal{N}}$ respectively. As following

$$\widetilde{\mathcal{N}} := \sin[\mathcal{N}\mathcal{K}L] / \sin[\mathcal{K}L],$$
$$\widehat{\mathcal{N}} := \sin[(\mathcal{N} - 1)\mathcal{K}L] / \sin[\mathcal{N}\mathcal{K}L], \tag{19}$$

with $\mathcal{N}$ corresponds to $a_f$, $P_1$ and $P_2$, respectively. It should be noted that $L$ denotes the length of the primitive cell associated with the Bloch wave vector $\mathcal{K}$ under consideration. Specifically, for $\mathcal{K}_1$ one has $L_1 = \lambda_0$, while for $\mathcal{K}_2$ one has $L_2 = (a_f + a_v)\lambda_0$. Especially, part I of the defective atomic lattice contains many internally embedded periodic structures, thus $\widehat{P_1} = \sin[(P_1 - 1)\mathcal{K}_2(a_f + a_v)\lambda_0] / (\widetilde{a}_f \cdot \sin[P_1\mathcal{K}_2(a_f + a_v)\lambda_0])$. By substituting Eq. (18) into Eq. (13), the matrix $M'$ expressed in terms of $\mathcal{K}_1$ and $\mathcal{K}_2$ can be obtained. Then, the left/right reflection coefficients are

$$r^l = -r^l_{\text{I}} \cdot \left[1 + 1/\left\{\widetilde{\mathcal{I}}\left(\widetilde{a}_f \cdot \widetilde{P}_1\right)^2\right\}\right],$$
$$r^r = r^r_{\text{II}} \cdot \left[1 + 1/\left\{\widetilde{\mathcal{I}}\left(\widetilde{P}_2\right)^2\right\}\right]. \tag{20}$$

Where $r^l_{\text{I}}(r^r_{\text{II}})$ is the reflection coefficient of Part I(II) for the probe field incident from the left (right)-side, namely

$$r^l_{\text{I}} = -\frac{M'_f(2,1)}{\left[M'_f(2,2) - \widehat{a_f}\right] - \widehat{P_1} e^{ika_v\lambda_0}},$$
$$r^r_{\text{II}} = \frac{M'_f(1,2)}{M'_f(2,2) - \widehat{P_2}}, \tag{21}$$

and $\widetilde{\mathcal{I}}$ is the synergy factor between the two parts of defective atomic lattice, given by

$$\widetilde{\mathcal{I}} = \left[\left(M'_f(2,2) - \widehat{a_f}\right)e^{-ika_v\lambda_0} - \widehat{P_1}\right] \\ \times \left(M'_f(2,2) - \widehat{P_2}\right)e^{-ika_v\lambda_0} + M'_f(1,2)M'_f(2,1). \quad (22)$$

It is thus clear that the left and right reflection coefficients obtained from the analytical transfer matrix $M'$ can clearly demonstrate the specific relationship between them and the lattice structure, which allows for more accurately and targetedly adjusting the parameters of the system and better understanding the essence of physical phenomena.

Similarly, the scattering matrix $S$ relates the outgoing (electric) field amplitudes ($E^r_{pr}$ and $E^r_{pl}$) to the incoming (electric) field amplitudes ($E^i_{pl}$ and $E^i_{pr}$) [see Fig. 1(c)]

$$\begin{bmatrix} E^r_{pr} \\ E^r_{pl} \end{bmatrix} = S \begin{bmatrix} E^i_{pl} \\ E^i_{pr} \end{bmatrix} = \begin{bmatrix} t & r^r \\ r^l & t \end{bmatrix} \begin{bmatrix} E^i_{pl} \\ E^i_{pr} \end{bmatrix}, \quad (23)$$

to verify the existence of lasing, we use the eigenvalues $\lambda_\pm$ of scattering matrix $S$ to explore the spectral singularity (SS):

$$\lambda_\pm = t \pm \sqrt{r^l r^r}, \quad (24)$$

the SS implies laser divergence, corresponding to $\lambda_+ \to \infty$, $\lambda_-^{-1} \to M(1,1)/2$ [46]. To make it easier to identify, we use the inverse of scattering matrix $S$, and its eigenvalues $\lambda_\pm^{-1}$ are expressed in terms of the elements of the transfer matrix, namely

$$\lambda_\pm^{-1} = \frac{M(2,2)}{1 \pm \sqrt{1 - M(1,1)M(2,2)}}. \quad (25)$$

Obviously, the SS corresponds to $\lambda_+^{-1} \to 0$ and $\lambda_-^{-1} \to 2/M(1,1)$. It is worth noting that, when the system exhibits the phenomenon of unidirectional reflection, the above eigenvalues become degenerate and equal to transmission coefficient of the system, corresponding to NHD ($\lambda_+^{-1} = \lambda_-^{-1}$). Intriguingly, in this paper, we further explore the lasing characteristic of unidirectional reflection, which requires $\lambda_+^{-1} = \lambda_-^{-1} \to 0$, it means that the URL corresponds to NHDSS point.

### III. NUMERICAL RESULTS AND DISCUSSIONS

#### A. The modulation of URL Corresponding to NHDSS point

In this section, we aim to explore how to achieve the unidirectional reflection lasing (URL) and analyze the physical essence. First, in the case of $\vartheta/\pi = 0$ (the maximum VIC), that the probe light is significantly amplified due to the strong quantum coherence effect caused by VIC [see Fig. 1(b)], we check the left-right reflections and the eigenvalues of the inverse of $S$ matrix v.s. probe detuning with modulating other parameters as shown in Figs. 2. It can be clearly seen that the peak of the left reflection (left peak) and the valley of the right reflection (right valley) are staggered [see Fig. 2(a) and 2(b)], corresponding to the spectral singularity (SS) ($\lambda_+^{-1} \to 0$) and the non-Hermitian degeneracy (NHD) point ($\lambda_+^{-1} \simeq \lambda_-^{-1}$), respectively [see Fig. 2(c) and 2(d)]. It is worth emphasizing that by adjusting the detuning of the coupling field (from resonance to off-resonance, of course, other parameters will also change accordingly.), the SS and the NHD point are approaching [compared Fig. 2(a) and 2(b) or Fig. 2(c) and 2(d)]. If the two points coincide, it corresponding to the URL. Next, we will try to achieve both the SS and the NHD in the same probe frequency.

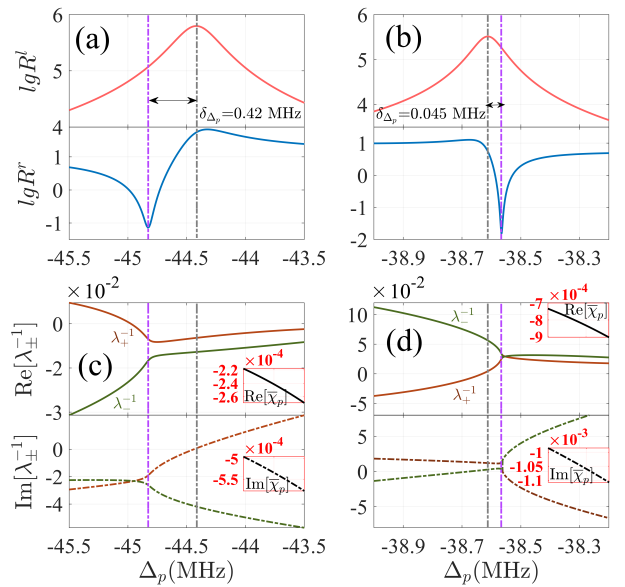

FIG. 2: (a)&(b) The logarithm of reflectivities on both sides of the probe $lgR^l$ and $lgR^r$ v.s. probe detuning $\Delta_p$. (c)&(d) The real and imaginary part of eigenvalues $\lambda_\pm^{-1}$ of the inverse of $S$ matrix v.s. probe detuning $\Delta_p$, where the inset figure plot the real and imaginary part of average probe susceptibility, $\mathrm{Re}[\bar{\chi}_p]$ and $\mathrm{Im}[\bar{\chi}_p]$. Here, (a)&(c) within $\Delta_c = 0$, $P_1 = 35$, $P_2 = 1200$, and (c)&(d) within $\Delta_c = -5$ MHz, $P_1 = 15$, $P_2 = 600$, and the same parameters are $\Omega_{c_0} = 30$ MHz, $\vartheta/\pi = 0$, $\phi_F/\pi = 0$, $a_f = 15$, $a_v = 150$, other relevant parameters for all panels are the same as Fig. 1.

In the following, we focus on the modulation of dipole-moment angle $\vartheta$ which can explore the effect of vacuum induced coherence $\Gamma_{32}$, try to find the NHDSS point. As shown in Fig. 3(a), as the angle $\vartheta$ changes, the frequency difference $\delta_{\Delta_p}$ between the left peak and the right valley also changes, [but it does not show a linear relationship, which can be seen from Eq. (6).]. Correspondingly, other parameters also need to be reset, such as the coupling

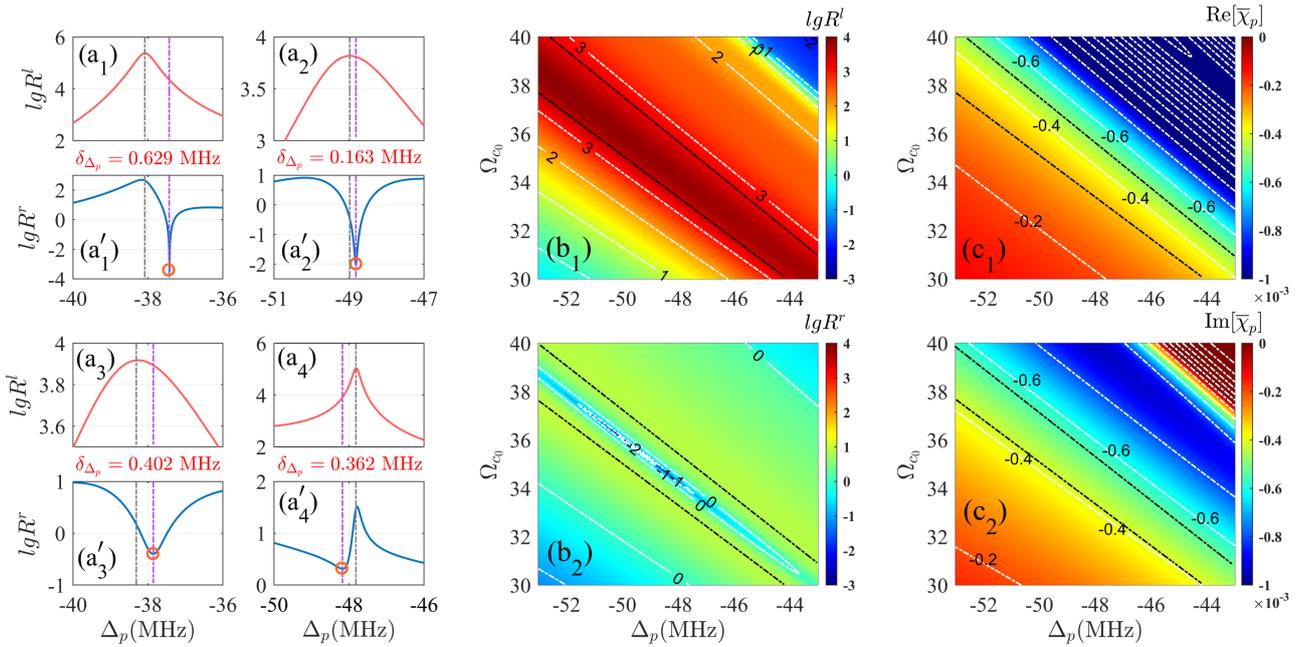

FIG. 3: (a) The logarithm of reflectivities on both sides of the probe $lgR^l$ and $lgR^r$ v.s. probe detuning $\Delta_p$ in difference parameter, whereas (a$_1$)&(a$_1'$) with $\vartheta/\pi = 0.1$, $\Omega_{c_0} = 25$ MHz, $\Delta_c = -10$ MHz, $P_1 = 36$; (a$_2$)&(a$_2'$) with $\vartheta/\pi = 0.2$, $\Omega_{c_0} = 35$ MHz, $\Delta_c = -10$ MHz, $P_1 = 41$; (a$_3$)&(a$_3'$) with $\vartheta/\pi = 0.3$, $\Omega_{c_0} = 30$ MHz, $\Delta_c = 10$ MHz, $P_1 = 41$; (a$_4$)&(a$_4'$) with $\vartheta/\pi = 0.4$, $\Omega_{c_0} = 50$ MHz, $\Delta_c = 20$ MHz, $P_1 = 65$. (b) The logarithm of reflectivities on both sides of the probe $lgR^l$ and $lgR^r$, and (c) The real and imaginary part of average probe susceptibility $\text{Re}[\overline{\chi}_p]$ and $\text{Im}[\overline{\chi}_p]$, both v.s. $\Delta_p$ and $\Omega_{c_0}$, the parameters of both are the same as (a$_2$)&(a$_2'$). The common parameters of all figures are $\phi_F/\pi = 0$, $a_f = 15$, $a_v = 150$, $P_2 = 800$. The other relevant parameters for all panels are the same as Fig. 2.

Rabi frequency and the lattice structure. This means that VIC is the key to amplify probe light, but it is not the only factor in forming unidirectional lasing. Specifically, the probe field can only be well amplified when it satisfies two-photon resonance [see Fig. 3(b) two-photon resonance energy levels], for $\Gamma_{32}$ is determined by $\vartheta$, it is not to see that the frequency region can be modulated by the angle $\vartheta$. It is worth emphasizing that when $\vartheta = 0$, the left peak achieves the best amplification, and the right vally reaches the deepest; when $\vartheta = 0.1$, the difference between the left peak and the right valley is the smallest, which is closest to the NHDSS point; when $\vartheta$ is further increased, the narrowband property and unidirectionality of the lasing will weaken or even disappear. In further, with the same angle $\vartheta$, the discontinuous NHDSS points (left peak corresponds to right valley) can be realized by modulating coupling Rabi frequency [see Figs. 3(b1) and 3(b2)]. For a more convenient description of the probe susceptibility, we replace the Gaussian-distributed probe susceptibility $\chi_p(x)$ with an average probe susceptibility $\overline{\chi}_p$. That is, the susceptibility of single filled lattice of length $\lambda_0$ no longer varies according to a Gaussian distribution along the position, but is instead represented by a rectangular, spatially averaged susceptibility. Correspondingly, the average probe susceptibility corresponding to NHDSS point also varies while remaining at a stable value $\overline{\chi}_{p0}$ ($\text{Re}[\overline{\chi}_{p0}] \approx -0.4 \times 10^{-3}$ and $\text{Im}[\overline{\chi}_{p0}] \approx -0.5 \times 10^{-3}$) shown in Figs. 3(c1) and 3(c2). The above reveal is related to the Bragg condition [42], given by

$$\frac{\omega_p}{c}\lambda_0 \left[P_1(a_f\overline{n_f(x)} + a_v) + P_2\overline{n_f(x)}\right] = \mathbb{N}\pi, \qquad (26)$$

where $c$ is the speed of light in vacuum, $\mathbb{N}$ is an arbitrary positive integer, and $\overline{n_f(x)}$ is the average refractive index of a filled lattice cell. Obviously, when NHDSS point appears, its corresponding average probe susceptibility $\overline{\chi}_p$ stabilizes at a certain value $\overline{\chi}_{p0}$. As the position of $\overline{\chi}_{p0}$ shifts, the position of NHDSS point also shifts, ensuring that the Bragg condition remains satisfied.

In the following, we will consider the modulation of total closed-loop phase $\phi_F$ on NHDSS point under the determined $\vartheta$ ($\vartheta/\pi = 0.125$), then plot the logarithm of reflectivities $lgR^l$ and $lgR^r$ with different phase $\phi_F$ in Figs. 4. Obviously, when $\phi_F/\pi = \pm 0.5$, $\delta_{\Delta_p}$ has almost dropped to 0 ($10^{-3}$ MHz $\sim 10^{-2}$ MHz), corresponding to a very well URL (with the same lattice structure); even more gratifyingly, when $\phi_F/\pi = -0.25$, $\delta_{\Delta_p} = 0$, this achieves a true NHDSS point, which corresponds a completely monochromatic URL achieving the resembling "unidirectional zero-width resonances"[46]. When the angle $\phi_F/\pi = 0.25$, the frequency difference between left peak and right valley is very large ($\delta_{\Delta_p} = 1.705$ MHz), and the left reflection bandwidth has exceeded 5 MHz, no longer possessing lasing characteristics. Even if it has the same lattice structure as $\phi_F/\pi = -0.25$. However, the NHDSS point can also be found in this phase, by changing the lattice structure and adjusting the appropriate coupling Rabi frequency and detuning, but the lasing intensity is somewhat reduced [see the illustration in Figs. 4(c1) and 4(c2) ]. It is obvious that the emergence of NHDSS point or the realization of URL is constrained by the lattice structure and coupling field. Next, we will discuss their relationship and physical essence in detail.

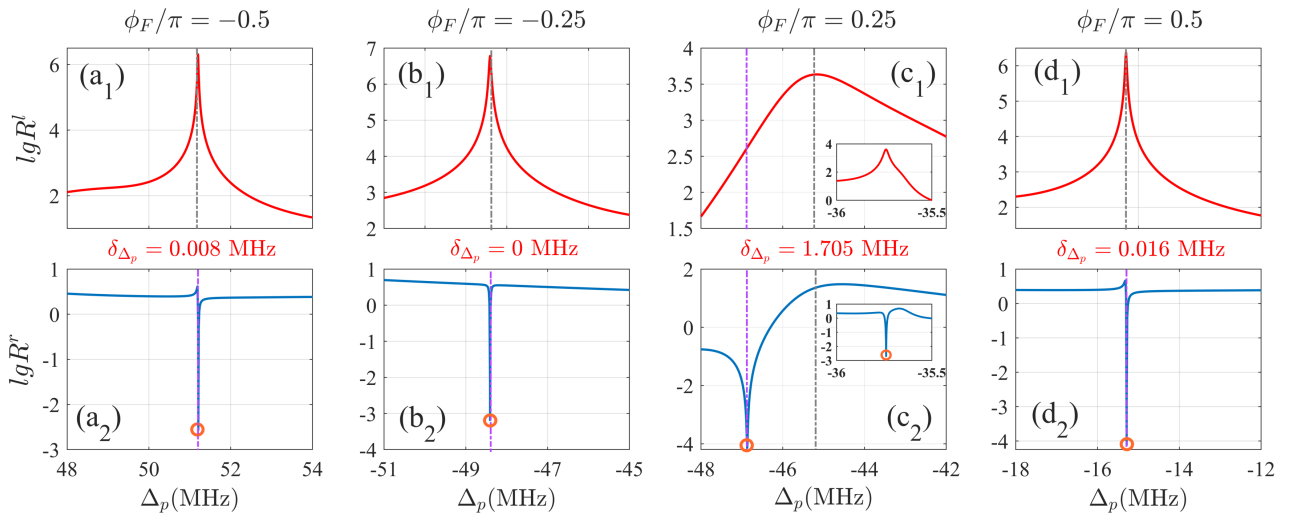

FIG. 4: (a)~(d) The logarithm of reflectivities on both sides of the probe $lgR^l$ and $lgR^r$ v.s. probe detuning $\Delta_p$ in difference parameter, whereas (a$_1$)&(a$_2$) with $\phi_F/\pi = -0.5$, $\Omega_{c_0} = 46$ MHz, $\Delta_c = -6.5$ MHz, $a_f = 30$, $a_v = 200$; (b$_1$)&(b$_2$) with $\phi_F/\pi = -0.25$, $\Omega_{c_0} = 38.2$ MHz, $\Delta_c = 10$ MHz, $a_f = 20$, $a_v = 200$; (c$_1$)&(c$_2$) with $\phi_F/\pi = 0.25$, $\Omega_{c_0} = 30$ MHz, $\Delta_c = -15$ MHz, $a_f = 20$, $a_v = 200$ (the inset of (c$_1$)&(c$_2$) with $\delta_{\Delta_p} = 0.0007$ MHz, $\phi_F/\pi = 0.25$, $\Omega_{c_0} = 40.5$ MHz, $\Delta_c = 5$ MHz, $a_f = 20$, $a_v = 180$); (d$_1$)&(d$_2$) with $\phi_F/\pi = 0.5$, $\Omega_{c_0} = 35.5$ MHz, $\Delta_c = 18.4$ MHz, $a_f = 30$, $a_v = 200$, and the common parameters of all figures are $\vartheta/\pi = 0.125$, $P_1 = 41$, $P_2 = 1200$. The other relevant parameters for all panels are the same as Fig. 2. Notably, the $\text{Im}[\overline{\chi}_{p0}]$ of (a)~(d) corresponding to NHDSS are (a) $-0.24 \times 10^{-3}$, (b) $-0.38 \times 10^{-3}$, (c) $-0.33 \times 10^{-3}$ (inset: $-0.16 \times 10^{-3}$), and (d) $-0.24 \times 10^{-3}$. The $\text{Re}[\overline{\chi}_{p0}]$ of (a)~(d) corresponding to NHDSS are (a) $-0.8 \times 10^{-3}$, (b) $-0.78 \times 10^{-3}$, (c) $0.02 \times 10^{-3}$ (inset: $-3 \times 10^{-3}$), and (d) $-0.8 \times 10^{-3}$.

### B. The verification of NHDSS by transfer matrix and probe susceptibility

In Sec. III A, we found that when changing dipole-moment angle $\vartheta$ or total closed-loop phase $\phi_F$, other parameters also need to be adjusted to find the NHDSS point. From Eq. (24) and Eq. (25), it can be seen that NHD requires $M(1,2) \to 0$, and SS requires $M(2,2) \to 0$. Thus, the transfer matrix elements $M(1,2)$ and $M(2,2)$ are the keys to realizing NHDSS. Next, we attempt to establish equations for the transfer matrix elements and the parameters of lattice structure/coupling field e.g., $a_f$, $a_v$, $P_1$, $P_2$, $\Omega_{c_0}$ and $\Delta_c$. Nevertheless, the transfer matrix of our system is unable to utterly satisfy but only approx-imate this condition. In this case, the applicability of $S$ matrix is limited, underscoring the critical necessity of seeking analytical threshold condition for NHDSS. Obviously, deriving the whole transfer matrix to investigate the correlation among parameters is profoundly arduous and complex. Using a global-to-local approach according to Eq. (18) to establish the correlation is simpler and more intuitive [56]. By applying the method of limits, we assume that our system is in full accordance with the conditions for realizing NHDSS, so that $M(1,2) = 0$ and $M(2,2) = 0$ can be used as boundary conditions and substituted into $M'$ matrix of whole medium to obtain

$$M'_f(1,1)\, e^{ik a_v \lambda_0} + M'_f(2,2)\, e^{-ik a_v \lambda_0} \\ = \widehat{P_1} + \widehat{a_f}\, e^{-ik a_v \lambda_0} + \widehat{P_2}\, e^{ik a_v \lambda_0}, \tag{27}$$

whereas $M'_f(2,2)$ of one filled cell satisfies specific lattice structural parameters, expressed in terms of $\mathcal{K}_1$ and $\mathcal{K}_2$, given by

$$M'_f(2,2) = \frac{e^{ik a_v \lambda_0} - \widehat{P_2}\,\widehat{a_f} e^{-ik a_v \lambda_0} - \widehat{P_2}\,\widehat{P_1}}{2i \widehat{P_2} \sin[k a_v \lambda_0]}, \tag{28}$$

which is determined by the parameters of lattice structure and coupling field. Therefore, as long as the parameters ensure that $M'_f(2,2)$ satisfies the Eq. (28), the NHDSS point will definitely appear. We can thus refer to this equation as the threshold condition for the NHDSS point. As shown in Figs. 5(a) and (b), $M'_f(2,2)$ intersects with $M_f(2,2)$ at $\Delta_p = -48.408$ MHz using the same parameters as Figs. 4(b). This means that $\Delta_p = -48.408$ MHz corresponds to an NHDSS point, which is completely consistent with Figs. 4(b), demonstrating the applicability of Eq. (28) in identifying NHDSS and the feasibility of the analytical threshold condition.

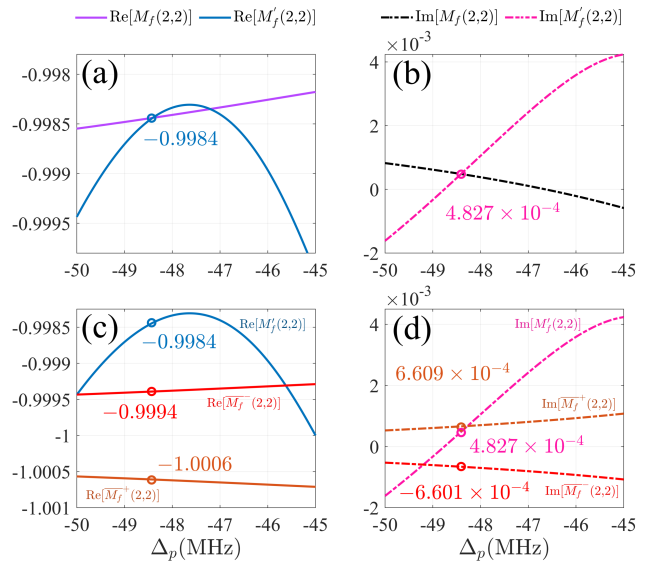

FIG. 5: (a)&(b) The $M_f(2,2)$ of one filled cell and the analytical threshold condition $M'_f(2,2)$, both adopts the numerical solution of the Gaussian-distributed susceptibility $\chi_p(x)$ v.s. $\Delta_p$. (c)&(d) The $\overline{M_f}^{\pm}(2,2)$ adopts the average susceptibility $\overline{\chi}_p$, and the $M'_f(2,2)$ of $\chi_p(x)$ both v.s. $\Delta_p$. The parameters are the same as those used for the NHDSS in Fig. 4(b).

Based on optical principles [57], we adopt the average probe susceptibility $\overline{\chi}_p$ to rectangularize the Gaussian distribution of one filled cell. Assuming that a thin vacuum layer exists on both sides of each filled cell, we can easily obtain $\overline{M_f}(2,2)$ in terms of $\overline{\chi}_p$, as

$$\overline{M_f}(2,2) = \exp\left[\pm i\pi \cos\theta \sqrt{1+\overline{\chi}_p}\right]. \tag{29}$$

There is bound to be a certain error between $\overline{M_f}(2,2)$ and $M'_f(2,2)$, and this error needs to be discussed. As shown in Figs. 5(c) and (d), $\overline{M_f}(2,2)$ and $M'_f(2,2)$ exhibit an acceptable minor deviation ($\simeq 10^{-3} \longrightarrow 0$), which result from the distinction between the average susceptibility $\overline{\chi}_p$ and the Gaussian-distributed susceptibility $\chi_p(x)$. Next, we set $\overline{M_f}(2,2) \approx M'_f(2,2)$ to verify the reveal mentioned above, without loss of accuracy and obtain

$$\mp \sqrt{1+\overline{\chi}_p} = \frac{i}{\pi \cos\theta} \ln\left[\overline{M_f}(2,2)\right] \approx \frac{i}{\pi \cos\theta} \ln\left[M'_f(2,2)\right], \tag{30}$$

with

$$\ln\left[M'_f(2,2)\right] \approx \mathring{\eta} \pm i(\pi - \mathring{\iota});\ \mathring{\eta} \approx \ln\left|\mathrm{Re}\left[M'_f(2,2)\right]\right|, \\ \pm i(\pi - \mathring{\iota}) \approx \ln\left(i\,\mathrm{Im}\left[M'_f(2,2)\right]/\left|\mathrm{Re}\left[M'_f(2,2)\right]\right| - 1\right), \tag{31}$$

here $\mathring{\eta}$ and $\mathring{\iota}$ are both small quantities, specifically corresponding to $-1.6 \times 10^{-3}$ and $5 \times 10^{-4}$ in Fig. 4(b), respectively. Notably, if $\mathrm{Im}[M'_f(2,2)]$ takes a positive or negative values, then we respectively obtain $+(\pi - \mathring{\iota})$ and $-(\pi - \mathring{\iota})$. Moreover, we obtain the relationship between the small quantities $\mathring{\eta}$ and $\mathring{\iota}$, given by

$$e^{\mp i\mathring{\iota}} + i\,\mathrm{Im}\left[M'_f(2,2)\right]e^{-\mathring{\eta}} = 1. \tag{32}$$

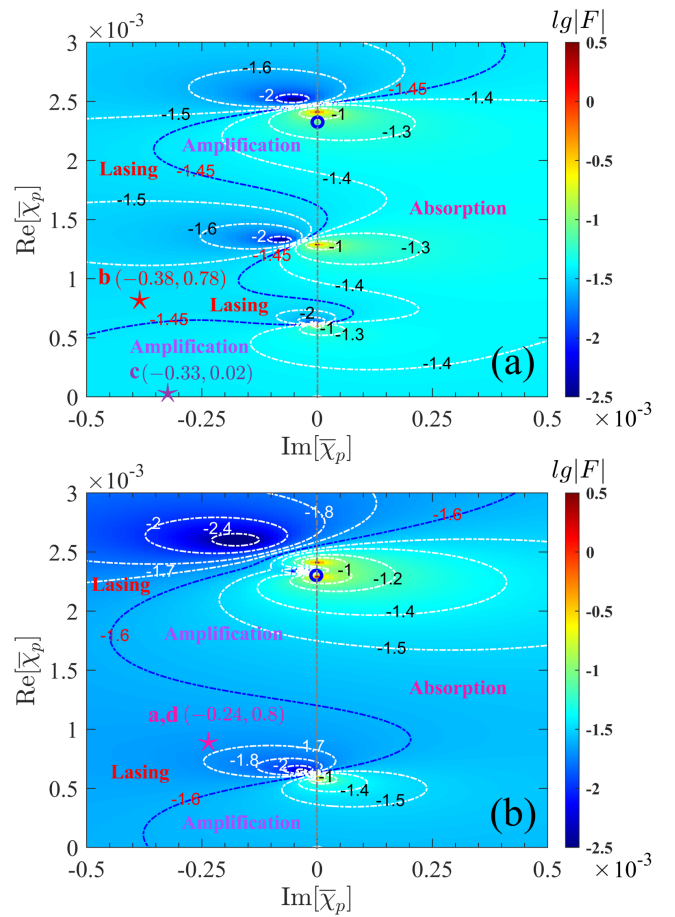

FIG. 6: (a)&(b) The logarithm of transcendental function $lg|F|$ both $v.s.$ $\mathrm{Re}[\overline{\chi}_p]$ and $\mathrm{Im}[\overline{\chi}_p]$, where we adopt the lattice structural parameters of Fig. 4(a)&(d) in (a), and follow those of Fig. 4(b)&(c) in (b). The $\mathrm{Re}[\overline{\chi}_{p0}]$ and $\mathrm{Im}[\overline{\chi}_{p0}]$ corresponding to Fig. 4 are marked by pentagrams, where $\mathrm{Re}[\overline{\chi}_{p0}]$ taken as its absolute value. The blue circle represents the "geometric" Bragg detuning $\mathfrak{z}$, and gray dash line represent the EIT regime.

By solving Eqs. (29) ∼ (30), we have

$$\begin{aligned}\operatorname{Re}\left[\overline{\chi}_p\right] &\approx \frac{2+\mathfrak{z}}{\zeta}\left(1-\frac{\mathring{\iota}}{\pi}\right)-2, \\ \operatorname{Im}\left[\overline{\chi}_p\right] &\approx \frac{2+\mathfrak{z}}{\zeta}\cdot\frac{\mathring{\eta}}{\pi},\end{aligned} \quad (33)$$

here $\mathfrak{z} = -2\Delta\lambda_{Lat}/\lambda_{Lat}$ and $\zeta = \lambda_p/\lambda_{Lat}$, both originating from the Bragg angle $\cos\theta$, thus we have defined $\mathfrak{z}$ as the "geometric" Bragg detuning according to Ref. [40, 53, 55]. obviously, the average probe susceptibility $\overline{\chi}_p$ corresponding to NHDSS still needs to meet the "geometric" Bragg detuning $\mathfrak{z}$. This is why the well URL corresponding to NHDSS point is unable to appear at the position of maximum probe gain. According to Eqs. (32) ∼ (33), we can easily derive the condition that generates the photonic-bandgap in a perfect atomic lattice, that $\operatorname{Re}\left[\overline{\chi}_p\right] \approx \mathfrak{z}$ and $\operatorname{Im}\left[\overline{\chi}_p\right] \approx 0$, when $\mathring{\eta}=0$, $\mathring{\iota}=0$ and $\operatorname{Im}[M'_f(2,2)] = 0$, which is consistent with Refs. [53, 55].

However, the small quantities $\mathring{\eta}$ and $\mathring{\iota}$ are no longer zero in the gain atomic systems, results in the range of the analytical threshold condition $M'_f(2,2)$ for the appearance of NHDSS are

$$\begin{aligned}\operatorname{Re}\left[M'_f(2,2)\right] &\in [-1+\mathring{\eta}, -1)\cup(-1, -1-\mathring{\eta}], \\ \operatorname{Im}\left[M'_f(2,2)\right] &\in [\mathfrak{z}\pi/2-\mathring{\iota}, 0)\cup(0, -\mathfrak{z}\pi/2+\mathring{\iota}].\end{aligned} \quad (34)$$

Notably, the small quantities $\mathring{\eta}$ and $\mathring{\iota}$ can be uniquely determined by the fixed lattice structural parameters. Thus, the probe susceptibility $\chi_p$ is modified, by tuning the coupling-field parameters $\Omega_{c_0}$ or $\Delta_c$, and corresponding to the shift NHDSS point. It is worth emphasizing that the small quantities $\mathring{\eta}$ and $\mathring{\iota}$ exhibit an extremely small variation (on the order of $10^{-6}$) limited by the order of average probe susceptibility $\overline{\chi}_p$. Combining Eqs. (28) to (30), we can obtain the transcendental function for the $\overline{\chi}_p$ of NHDSS point:

$$\begin{aligned}F\left(\overline{\chi}_p\right) := &\, 2i\sin\left[a_v\pi(1+\mathcal{B})\right]e^{\pm i\pi\zeta(1-\mathcal{B})\sqrt{1+\overline{\chi}_p}} - \frac{\sin\left[P_2\cos^{-1}\left\{\frac{1}{2}\pi^2\mathcal{B}\overline{\chi}_p-1\right\}\right]}{\sin\left[(P_2-1)\cos^{-1}\left\{\frac{1}{2}\pi^2\mathcal{B}\overline{\chi}_p-1\right\}\right]}e^{ia_v\pi(1+\mathcal{B})} \\ &+ \frac{\sin\left[(a_f-1)\cos^{-1}\left\{\frac{1}{2}\pi^2\mathcal{B}\overline{\chi}_p-1\right\}\right]}{\sin\left[a_f\cos^{-1}\left\{\frac{1}{2}\pi^2\mathcal{B}\overline{\chi}_p-1\right\}\right]}e^{-ia_v\pi(1+\mathcal{B})} + \frac{\sin\left[\cos^{-1}\left\{\frac{1}{2}\pi^2\mathcal{B}\overline{\chi}_p-1\right\}\right]}{\sin\left[a_f\cos^{-1}\left\{\frac{1}{2}\pi^2\mathcal{B}\overline{\chi}_p-1\right\}\right]} \\ &\times \frac{\sin\left[(P_1-1)\cos^{-1}\left\{\cos[a_v\pi]\cdot\left(\frac{\sin[a_f\cos^{-1}\{\frac{1}{2}\pi^2\mathcal{B}\overline{\chi}_p-1\}]}{\sin[\cos^{-1}\{\frac{1}{2}\pi^2\mathcal{B}\overline{\chi}_p-1\}]}\left(\frac{1}{2}\pi^2\mathcal{B}\overline{\chi}_p-1\right)-\frac{\sin[(a_f-1)\cos^{-1}\{\frac{1}{2}\pi^2\mathcal{B}\overline{\chi}_p-1\}]}{\sin[\cos^{-1}\{\frac{1}{2}\pi^2\mathcal{B}\overline{\chi}_p-1\}]}\right)\right\}\right]}{\sin\left[P_1\cos^{-1}\left\{\cos[a_v\pi]\cdot\left(\frac{\sin[a_f\cos^{-1}\{\frac{1}{2}\pi^2\mathcal{B}\overline{\chi}_p-1\}]}{\sin[\cos^{-1}\{\frac{1}{2}\pi^2\mathcal{B}\overline{\chi}_p-1\}]}\left(\frac{1}{2}\pi^2\mathcal{B}\overline{\chi}_p-1\right)-\frac{\sin[(a_f-1)\cos^{-1}\{\frac{1}{2}\pi^2\mathcal{B}\overline{\chi}_p-1\}]}{\sin[\cos^{-1}\{\frac{1}{2}\pi^2\mathcal{B}\overline{\chi}_p-1\}]}\right)\right\}\right]}.\end{aligned} \quad (35)$$

where we define $\mathcal{B} = -\mathfrak{z}/(2+\mathfrak{z})$ as Bragg condition parameter. It can be seen from Eq. (35) that the average probe susceptibility making $F \approx 0$ (Alternatively $F \to 0$) both corresponds to the NHDSS point and satisfies the Bragg condition, perfectly corresponding to the URL, based on the fixed lattice structural parameters.

In Figs. 6, there are three regions: (I) the absorption region that $\operatorname{Im}[\overline{\chi}_p] \geq 0$; (II) the amplification region that $\operatorname{Im}[\overline{\chi}_p] < 0$ and $F$ is more than threshold of lasing ($F$ lies further from 0 and tends toward 1); (III) the lasing region that $\operatorname{Im}[\overline{\chi}_p] < 0$ and $F$ is less than threshold of lasing ($F$ lies closer to 0). It is worth emphasizing that the thresholds vary under different parameters, which can be seen from the blue dot-dashed lines in Figs. 6(a) and 6(b), respectively. Here we refer to them as threshold lines, because these two contour lines exactly include all points with $F \to 0$ and exhibit the characteristics of spectral singularities. Next, we will deeply analysis the reflection characteristics corresponding to these three regions. First, in region (I), the photonic bandgap must locate in the EIT window ($\operatorname{Im}[\overline{\chi}_p]=0$) with the Bragg condition that $\operatorname{Re}[\overline{\chi}_p] = \mathfrak{z}$ [see blue circles in Figs. 6(a) and 6(b) corresponding to $lg|F| \approx -1$], which is well consistent with the perfect atomic lattice system in Ref. [55]. Especially, the unidirectional reflection band can also occurs when $\operatorname{Re}[\overline{\chi}_p] = \mathfrak{z}/2$ or $\mathfrak{z}/4$ (which also correspond to $lg|F| \approx -1$) in a defective atomic lattice, expect $\operatorname{Re}[\overline{\chi}_p] = \mathfrak{z}$. Then, In region (II), since the probe susceptibility is negative, the probe light can be well amplified but cannot output lasing, and the susceptibility in Fig. 4(c) exactly locates in this region [see the point c in Fig. 6(a)]. Only when the value of $F$ breaks through the threshold line and the corresponding susceptibility must be in region (III) that the URL can be realized. Fig. 4(b), Figs. 4(a) and 4(d) correspond to point b in Fig. 6(a), and points a and d in Fig. 6(b), respectively. It should be emphasized that the URL does not correspond to the best $F \to 0$, which is mainly due to the small deviation caused by the approximation of Eq. (27) and Eq. (30). Figs. 6 highlight that the appearance of unidirectionality alone does not necessarily indicate URL, the decisive criterion is whether the corresponding eigenvalues $\lambda_\pm^{-1}$ is degenerate and approach zero, that is, whether it corresponds to a NHDSS point.

## IV. CONCLUSIONS

In summary, we investigate and discuss the tunable URL in the 1D defective atomic lattice. The atoms trapped in lattice cells are driven into three-level $V$-type active atomic system for amplifying the weak probe field via VIC, and the 1D defective atomic lattice is used to break the spatial symmetry and provide distributed feedback regime. The URL means that both unidirectional reflection and lasing output occur simultaneously, this requires that the inverse of the eigenvalues of the scattering matrix satisfies both $\lambda_+^{-1} \simeq \lambda_-^{-1}$ and $\lambda_+^{-1} \to 0$ which corresponding to the NHDSS point. Therefore, we first modulate the relevant parameters to search for NHDSS points, and the numerical results show that as long as any parameters among the lattice structure parameters, the Rabi frequency and detuning of the coupling field, the closed-loop phase $\phi_F$, and the angle of the dipole moments changes, all other physical quantities must be adjusted to make it possible to rediscover the NHDSS point. Then we further verify the URL and analyze its physical essence by solving the transcendental equation of susceptibility that satisfies NHDSS. Our study provides theoretical foundation for NHDSS in non-Hermitian systems, and unidirectional reflection lasing offers potential pathways for realizing non-reciprocal optical circuits and high-performance all-optical controlled unidirectional devices for integrated photonic circuits.

## V. ACKNOWLEDGMENTS

This work is supported by the Hainan Provincial Natural Science Foundation of China (Grant Nos. 121RC539) and the National Natural Science Foundation of China (Grant No. 12204137). This project is also supported by the specific research fund of The Innovation Platform for Academicians of Hainan Province Grant (No.YSPTZX202215).

[1] L. Zhang, L. Zhao, Y.-J. Miao, and J. Dong, Elliptically polarized, nanosecond dual-pulse Raman laser with tunable pulse interval and pulse amplitude ratio, Opt. Laser Technol. **171**, 110397 (2024).

[2] C. Wang, K.-X. Yang, L.-Y. Qu, Y.-D. Li, R.-C. Zhao, X.-H. Xue, T.-D. Chen, and X.-K. Dou, Coherent Two-Photon Atmospheric Lidar Based on Up-Conversion Quantum Erasure, ACS Photonics **11**, 3528-3535 (2024).

[3] L.-C. Liu, C. Wu, W. Li, Y.-A. Chen, X.-P. Shao, F. Wilczek, F.-H. Xu, Q. Zhang, and J.-W. Pan, Active Optical Intensity Interferometry, Phys. Rev. Lett. **134**, 180201 (2025).

[4] R.A. Hanoon, A.H. Abdulhadi, and A.K. Abass, Dual-wavelength mode-locked erbium fiber laser utilizing a Ge-PCF saturable absorber, Appl. Opt. **63**, 8124-8130 (2024).

[5] A.-K. Yang, Z.-Y. Li, M.P. Knudson, A.J. Hryn, W.-J. Wang, K. Aydin, and T.W. Odom, Unidirectional Lasing from Template-stripped Two-dimensional Plasmonic Crystals, ACS Nono **9**, 11582-11588 (2015).

[6] J. Dixon, M. Lawrence, D.R. Barton, and J. Dionne, Self-Isolated Raman Lasing with a Chiral Dielectric Metasurface, Phys. Rev. Lett. **126**, 123201 (2021).

[7] Q.-H. Song, L.-Y. Liu, S.-M. Xiao, X.-C. Zhou, W.-C. Wang, and L. Xu, Unidirectional High Intensity Narrow-Linewidth Lasing from a Planar Random Microcavity Laser, Phys. Rev. Lett. **96**, 033902 (2006).

[8] H. Ramezani, S. Kalish, I. Vitebskiy, and T. Kottos, Unidirectional Lasing Emerging from Frozen Light in Nonreciprocal Cavities, Phys. Rev. Lett. **112**, 043904 (2014).

[9] N.T. Otterstrom, R.O. Behunin, E.A. Kittlaus, Z. Wang, and P.T. Rakich, A silicon Brillouin laser, Science **360**, 1113 (2018).

[10] N.T. Otterstrom, E.A. Kittlaus, S. Gertler, R.O. Behunin, A.L. Lentine, and P.T. Rakich, Resonantly enhanced nonreciprocal silicon Brillouin amplifier, Optica **6**, 1117 (2019).

[11] B. Bahari, A. Ndao, F. Vallini, A.E. Amili, Y. Fainman, B. Kanté, Nonreciprocal lasing in topological cavities of arbitrary geometries, Science **358**, 636-640 (2017).

[12] A. Muñoz de las Heras and I. Carusotto, Unidirectional lasing in nonlinear Taiji microring resonators, Phys. Rev. A **104**, 043501 (2021).

[13] Z.-W. Dai, W.-B. Lin, and S. Iwamoto, Rate equation analysis for deterministic and unidirectional lasing in ring resonators with an S-shaped coupler, Jpn. J. Appl. Phys. **63**, 02SP54 (2024).

[14] C.-Y. Lei and J. Ren, Quantum-interference-enhanced phonon laser in cavity optomechanics, Phys. Rev. Appl. **19**, 054093 (2023).

[15] Y.-R. Zhou, Q.-F. Zhang, F.-F. Liu, Y.-H. Han, Y.-P. Gao, L. Fan, R. Zhang, and C. Cao, Quantum-interference-enhanced phonon laser in cavity optomechanics, Opt. Express **32**, 2786-2803 (2024).

[16] M. Wang, Y.-C. Lei, Z.-G. Hu, C.-H. Lao, Y.-L. Wang, X. Zhou, J.-C. Li, Q.-F. Yang, and B.-B. Li, Coupling ideality of standing-wave supermode microresonators, Photonics Res. **12**, 1610-1618 (2024).

[17] Y.-J. Xu, and J. Song, Nonreciprocal magnon laser, Opt. Lett. **46**, 5276-5279 (2021).

[18] K.-W. Huang, Y. Wu, and L.-G. Si, Parametric-amplification-induced nonreciprocal magnon laser, Opt. Lett. **47**, 3311-3314 (2022).

[19] X.-W. He, Z.-Y. Wang, X. Han, S. Zhang, and H.-F. Wang, Parametrically amplified nonreciprocal magnon laser in a hybrid cavity optomagnonical system, Opt. Express **31**, 43506-43517 (2023).

[20] M.H. Song, B. Park, S. Nishimura, T. Toyooka, I.J. Chung, Y. Takanishi, K. Ishikawa, and H. Takezoe, Electrotunable Non-reciprocal Laser Emission from a Liquid-Crystal Photonic Device, Adv. Funct. Mater. **16**, 1793-1798 (2006).

[21] S.-Y. Ji, Y.-F. Zhou, L. Xiong, X.-Y. Liu, T. Zhu, X.-Q. Zhan, Y.-L. Yan, J.-N. Yao, K. Wang, and Y.-S. Zhao,

Nonreciprocal Circularly Polarized Lasing from Organic Achiral Microcrystals, J. Am. Chem. Soc. **147**, 16674-16680 (2025).

[22] J.M. Lee, S. Factor, Z. Lin, I. Vitebskiy, F.M. Ellis, and T. Kottos, Reconfigurable Directional Lasing Modes in Cavities with Generalized $\mathcal{PT}$-Symmetry, Phys. Rev. Lett. **112**, 253902 (2014).

[23] B. Megyeri, G. Harvie, A. Lampis, and J. Goldwin, Directional Bistability and Nonreciprocal Lasing with Cold Atoms in a Ring Cavity, Science **121**, 163603 (2018).

[24] X.-H. Wen, X.-H. Zhu, A. Fan, W.Y. Tam, J. Zhu, H.-W. Wu, F. Lemoult, M. Fink, and J. Li, Unidirectional amplification with acoustic non-Hermitian space-time varying metamaterial, Commun. Phys. **5**, 18 (2022).

[25] S. Landers, W. Tuxbury, I. Vitebskiy, T. Kottos, Unidirectional Amplification in the Frozen Mode Regime Enabled by a Nonlinear Defect, Opt. Lett. **49**, 4967-4970 (2024).

[26] Y. Geng, X.-S. Pei, G.-R. Li, X.-Y. Lin, H.-X. Zhang, D. Yan, and H. Yang, spatial susceptibility modulation and controlled unidirectional reflection amplification via four-wave mixing, Opt. Express **31**, 38228-38239 (2023).

[27] D. Song, X. Li, H.-T. Zhou, J.-J. Xue, R.-N. Li, D. Wang, B.-D. Yang, and J.-X. Zhang, Optical reciprocity-nonreciprocity-amplification conversion based on degenerate four-wave mixing, J. Opt. Soc. Am. B **41**, 984-991 (2024).

[28] Y. Geng, C. Peng, X.-F. Zheng, D. Yan, H.-X. Zhang, J.-H. Wu, and H. Yang, Unidirectional amplified reflections based on four-wave mixing and spontaneously generated coherence, Phys. Lett. A **534**, 130241 (2025).

[29] J.-H. Wu and J.-Y. Gao, Phase control of light amplification without inversion in a system with spontaneously generated coherence, Phys. Rev. A **65**, 063807 (2002).

[30] J.-H. Wu, H.-F. Zhang, and J.-Y. Gao, Probe gain with population inversion in a four-level atomic system with vacuum induced coherence, Opt. Lett. **28**, 654-656 (2003).

[31] Y. Zhang, Y.-M. Liu, T.-Y. Zheng, and J.-H. Wu, Light reflector, amplifier, and splitter based on gain-assisted photonic band gaps, Phys. Rev. A **94**, 013836 (2016).

[32] H.S. Han, A. Lee, K. Sinha, F.K. Fatemi, and S.L. Rolston, Observation of vacuum-induced collective quantum beats, Phys. Rev. Lett. **127**, 073604 (2021).

[33] M. Hennrich, A. Kuhn, and G. Rempe, Transition from Antibunching to Bunching in Cavity QED, Phys. Rev. Lett. **94**, 053604 (2005).

[34] Z.-M. Wu, J.-H. Li, and Y. Wu, Vacuum-induced quantum-beat-enabled photon anti-bunching, Phys. Rev. A **108**, 023727 (2023).

[35] Z.-M. Wu, J.-H. Li, and Y. Wu, Phase-engineered photon correlations in weakly coupled nanofiber cavity QED, Phys. Rev. A **109**, 033709 (2024).

[36] Z.-M. Wu, J.-H. Li, and Y. Wu, Coherent perfect absorption accompanying subnatural narrowing by vacuum-induced coupling, Phys. Rev. A **110**, 033722 (2024).

[37] D. A. Steck, Rubidium 85 D line data, http://steck.us/alkalidata (revision 2.3.3, 28 May 2024).

[38] G.-R. Li, Y. Geng, X.-S. Pei, J.-H. Wu, X.-Y. Lin, D. Yan, H.-X. Zhang, and H. Yang, Phase and detuning control of the unidirectional reflection amplification based on the broken spatial symmetry, Opt. Express **32**, 12839-12851 (2024).

[39] A. Schilke, C. Zimmermann, P.W. Courteille, and W. Guerin, Optical parametric oscillation with distributed feedback in cold atoms, Nat. Photonics **6**, 101-104 (2011).

[40] J.-H. Wu, M. Artoni, and G.-C. La Rocca, Two-color lasing in cold atoms, Phys. Rev. A **88**, 043823 (2013).

[41] T. Nadolny, M. Brunelli, and C. Bruder, Nonreciprocal Interactions Induce Frequency Shifts in Superradiant Lasers, Phys. Rev. Lett. **134**, 193603 (2025).

[42] T.-M. Li, H. Yang, M.-H. Wang, C.-P. Yin, T.-G. Zhang, and Y. Zhang, Unidirectional photonic reflector using a defective atomic lattice, Phys. Rev. Res. **6**, 023122 (2024).

[43] Q.-Y. Xu, G.-R. Li, Y.-T. Zheng, D. Yan, H.-X. Zhang, T.-G. Zhang, and H. Yang, Broadband unidirectional reflection amplification in a one-dimensional defective atomic lattice, Phys. Rev. A **110**, 063724 (2024).

[44] J.-H. Wu, M. Artoni, and G.-C. La Rocca, Non-hermitian degeneracies and unidirectional reflectionless atomic lattices, Phys. Rev. Lett. **113**, 123004 (2014).

[45] L.-J. Yuan, and Y.-Y. Lu, Unidirectional reflectionless transmission for two-dimensional PT-symmetric periodic structures, Phys. Rev. A **100**, 053805 (2019).

[46] A. Mostafazadeh, Spectral singularities of complex scattering potentials and infinite reflection and transmission coefficients at real energies, Phys. Rev. Lett. **102**, 220402 (2009).

[47] A. Mostafazadeh, Optical spectral singularities as threshold resonances, Phys. Rev. A **83**, 045801 (2011).

[48] A. Mostafazadeh, Semiclassical analysis of spectral singularities and their applications in optics, Phys. Rev. A **84**, 023809 (2011).

[49] A. Mostafazadeh, Nonlinear spectral singularities of a complex barrier potential and the lasing threshold condition, Phys. Rev. A **87**, 063838 (2013).

[50] H. Ramezani, H.-K. Li, Y. Wang, and X. Zhang, Unidirectional Spectral Singularities, Phys. Rev. Lett. **113**, 263905 (2014).

[51] M. Artoni, G. C. La Rocca, and F. Bassani, Resonantly absorbing one-dimensional photonic crystals, Phys. Rev. E **72**, 046604 (2005).

[52] J.-W. Gao, Y. Zhang, N. Ba, C.-L. Cui, and J.-H. Wu, Dynamically induced double photonic bandgaps in the presence of spontaneously generated coherence, Opt. Lett. **35**, 709-711 (2010).

[53] A. Schilke, C. Zimmermann, and W. Guerin, Photonic properties of one-dimensionally-ordered cold atomic vapors under conditions of electromagnetically induced transparency, Phys. Rev. A **86**, 023809 (2012).

[54] H. Yang, L. Yang, X.-C. Wang, C.-L. Cui, Y. Zhang, and J.-H. Wu, Dynamically controlled two-color photonic band gaps via balanced four-wave mixing in one-dimensional cold atomic lattices, Phys. Rev. A **88**, 063832 (2013).

[55] H. Yang, T.-G. Zhang, Y. Zhang, and J.-H. Wu, Dynamically tunable three-color reflections immune to disorder in optical lattices with trapped cold $^{87}$Rb atoms, Phys. Rev. A **101**, 053856 (2020).

[56] A. Mostafazadeh, Transfer matrix in scattering theory: A survey of basic properties and recent developments, Turk. J. Phys. **44**, 472-527 (2020).

[57] M. Born and E. Wolf, *Principles of Optics* (Cambridge University Press, Cambridge, UK, 1999).